\documentclass[english,aps,preprint]{revtex4}
\usepackage[T1]{fontenc}
\usepackage[latin9]{inputenc}
\usepackage{array}
\usepackage{float}
\usepackage{multirow}
\usepackage{amsmath}
\usepackage{amssymb}
\usepackage{graphicx}

\makeatletter

\providecommand{\tabularnewline}{\\}

\@ifundefined{textcolor}{}
{%
 \definecolor{BLACK}{gray}{0}
 \definecolor{WHITE}{gray}{1}
 \definecolor{RED}{rgb}{1,0,0}
 \definecolor{GREEN}{rgb}{0,1,0}
 \definecolor{BLUE}{rgb}{0,0,1}
 \definecolor{CYAN}{cmyk}{1,0,0,0}
 \definecolor{MAGENTA}{cmyk}{0,1,0,0}
 \definecolor{YELLOW}{cmyk}{0,0,1,0}
}

\makeatother

\usepackage{babel}
\begin{document}

\title{Excited neutrino search potential of the FCC-based electron-hadron
colliders}

\author{A. Caliskan}
\email{acaliskan@gumushane.edu.tr}

\affiliation{Gümü\c{s}hane University, Faculty of Engineering and Natural Sciences,
Department of Physics Engineering, 29100, Gümü\c{s}hane, Turkey}
\begin{abstract}
The production potential of the excited neutrinos at the FCC-based
electron-hadron colliders, namely the ERL60$\otimes$FCC with $\sqrt{s}=3.46$
TeV, the ILC$\otimes$FCC with $\sqrt{s}=10$ TeV, and the PWFA-LC$\otimes$FCC
with $\sqrt{s}=31.6$ TeV, has been analyzed. The branching ratios
of the excited neutrinos have been calculated for the different decay
channels and shown that the dominant channel is $\nu^{\star}\rightarrow eW^{+}$.
We have calculated the production cross sections with the process
of $ep\rightarrow\nu^{\star}q\rightarrow eW^{+}q$ and the decay widths
of the excited neutrinos with the process of $\nu^{\star}\rightarrow eW^{+}$.
The signals and corresponding backgrounds are studied in detail to
obtain accessible mass limits. It is shown that the discovery limits
obtained on the mass of the excited neutrino are $2452$ GeV for $L_{int}=100$
$fb^{-1}$, $5635$ GeV for $L_{int}=10$ $fb^{-1}$ ($6460$ GeV
for $L_{int}=100$ $fb^{-1}$), and $10200$ GeV for $L_{int}=1$
$fb^{-1}$ ($13960$ GeV for $L_{int}=10$ $fb^{-1}$), for the center-of-mass
energies of $3.46$, $10$, and $31.6$ TeV, respectively.
\end{abstract}
\maketitle

\section{\i ntroduction}

The Standard Model (SM) of the particle physics has so far been in
agreement with the results of numerous experiments. The discovery
of the Higgs boson \cite{ATLAS collaboration} has also increased
the reliability of the SM. However, there are some problems which
have not been entirely solved by the SM such as quark-lepton symmetry,
family replication, number of families, fermion's masses and mixing
pattern, hierarchy problems etc. A number of theories beyond the SM
(BSM), including extra dimensions, supersymmetry (SUSY), compositeness
and so on, have been proposed for solving these problems. One of the
most important of these theories is compositeness in which quarks
and leptons have a substructure called preon \cite{I.A. DSouza}.
The composite models have been characterized by an energy scale, namely
compositeness scale, $\Lambda$. A typical consequence of the compositeness
is the appearance of excited leptons and quarks \cite{J.H. K=0000FChn,U.Baur}.
Charged ($e^{\star}$, $\mu^{\star}$,$\tau^{\star}$) and neutral
($\nu_{e}^{\star}$, $\nu_{\mu}^{\star}$ , $\nu_{\tau}^{\star}$)
excited leptons are predicted by the composite models. The SM fermions
are considered as ground states of a rich and heavier spectrum of
the excited states. An excited spin-1/2 lepton is considered to be
the lowest radial and orbital excitation. Excited states with spin-3/2
are also expected to exist \cite{Y. Tosa}. 

No evidence for excited lepton production is found in studies using
data samples collected by the experiments, namely LEP \cite{LEP},
HERA \cite{HERA}, Tevatron \cite{Tevatron}, CMS \cite{CMS} and
ATLAS \cite{ATLAS} collaborations. For the excited electron \cite{O. Cakir single production,A.Ozansoy search},
muon \cite{A.Caliskan} and neutrino \cite{single production,3/2 neutrino,k=0000F6ksal,V.Ar=000131},
there are some phenomenological studies at the future high energy
colliders.

Current experimental lower bounds on the masses of the excited neutrinos
are $m_{\nu^{\star}}>102.6$ GeV \cite{LEP} from LEP - L3 collaboration
(pair production) assuming $f=-f'=1$, $m_{\nu^{\star}}>213$ GeV
\cite{particle sata group} at 95\% C.L. from HERA-H1 collaboration
(single production) assuming $f=f'=1$ and $m_{\nu^{\star}}>1.6$
TeV \cite{particle sata group}, namely the strongest limit, from
LHC-ATLAS collaboration (pair production) assuming $f=f'=1$.

The Future Circular Collider (FCC) is a post-LHC accelerator project
\cite{FCC web}, with $\sqrt{s}=100$ TeV, proposed at CERN and supported
by European Union within Horizon $2020$ Framework Programme for Research
and Innovation. Besides the $pp$ option, it includes $e^{+}e^{-}$
collider option (TLEP) at the same tunnel \cite{TLEP}. Construction
of the future $e^{+}e^{-}$ and $\mu^{+}\mu^{-}$ colliders tangential
to the FCC will also provide several $ep$ and $\mu p$ collider options
\cite{Sultansoy rome}.

In this paper we analyze the potential of the FCC-based $ep$ colliders,
namely ERL60$\otimes$FCC, ILC$\otimes$FCC and PWFA-LC$\otimes$FCC,
for the excited neutrino searches. The ERL60 denotes energy recovery
linac proposed for the LHeC main option \cite{LHeC web}, and can
also be used for the FCC-based $ep$ colliders. The ILC and the PWFA-LC
mean International Linear Collider \cite{ILC}, and Plasma Wake Field
Accelerator Linear Collider \cite{PWLC}, respectively. The FCC-based
ILC$\otimes$FCC and PWFA-LC$\otimes$FCC colliders have been proposed
in ref. \cite{YC Acar}. Center-of-mass energy and luminosity values
of the FCC-based $ep$ colliders are given in Table I \cite{YC Acar,FCC based}. 

We introduce the effective Lagrangian, the decay widths, and the branching
ratios of the excited neutrinos in Section II. In Section III, we
analyze the signal and backgrounds for the process $ep\rightarrow\nu^{\star}q\rightarrow eW^{+}q$
, and finally we summarize our results in Section IV.

\begin{table}

\caption{Main parameters of the FCC-based $ep$ colliders.}
\centering{}%
\begin{tabular}{|c|c|c|c|}
\hline 
Colliders  & $E_{e}$ (TeV) & CM Energy (TeV) & $L_{int}$ ($fb^{-1}$ per year)\tabularnewline
\hline 
\hline 
ERL60$\otimes$FCC & 0.06 & 3.46 & 100\tabularnewline
\hline 
ILC$\otimes$FCC & 0.5 & 10 & 10-100\tabularnewline
\hline 
PWFA-LC$\otimes$FCC & 5 & 31.6 & 1-10\tabularnewline
\hline 
\end{tabular}
\end{table}

\section{PRODUCTION OF THE EXCITED NEUTRINOS}

The interaction between a spin-1/2 excited lepton, a gauge boson ($V=\gamma,Z,W^{\pm}$)
and the ordinary leptons is described by $SU(2)\times U(1)$ invariant
Lagrangian \cite{U.Baur,Hagiwara,Boudjema} as

\begin{equation}
L=\frac{1}{2\Lambda}\overline{l_{R}^{*}}\sigma^{\mu\nu}\left[fg\frac{\vec{\tau}}{2}\centerdot\vec{W}_{\mu\nu}+f^{\prime}g^{\prime}\frac{Y}{2}B_{\mu\nu}\right]l_{L}+h.c.,
\end{equation}
where $\Lambda$ is the new physics scale that responsible for the
existence of the excited leptons; $W_{\mu\nu}$ and $B_{\mu\nu}$
are the field strength tensors, $g$ and $g^{\prime}$ are the gauge
couplings, $f$ and $f^{\prime}$ are the scaling factors for the
gauge couplings of $SU(2)$ and $U(1)$; $\sigma^{\mu\nu}=i(\gamma^{\mu}\gamma^{\nu}-\gamma^{\nu}\gamma^{\mu})/2$
where $\gamma^{\mu}$ are the Dirac matrices, $\overrightarrow{\tau}$
denotes the Pauli matrices, and Y is hypercharge.

The excited neutrinos have three decay modes, namely radiative decay
$\nu^{\star}\rightarrow\nu\gamma$, neutral weak decay $\nu^{\star}\rightarrow\nu Z$
, and charged weak decay $\nu^{\star}\rightarrow eW^{+}$. The branching
ratios (BR) of the excited neutrino for the coupling of $f=f^{\prime}=1$
and $f=-f^{\prime}=1$ are given in Fig. 1. One may note that the
electromagnetic interaction between the excited neutrino and ordinary
neutrino, namely $\gamma$ - channel, vanishes for the coupling of
$f=f^{\prime}=1$. It is clearly seen that the W-channel whose branching
ratio is $\sim$ 60\% become dominant. For the coupling of $f=-f^{\prime}=1$,
the branching ratio for the individual decay channels reaches to the
constant values 60\% for the W-channel, 12\% for the Z-channel, and
28\% for the $\gamma$ -channel at higher mass values. Since the charged
weak decay ($\nu^{\star}\rightarrow eW^{+}$) is dominant for both
cases, we preferred this channel for the investigation of the excited
neutrino in this paper.

Neglecting the SM lepton mass, we find the decay width of excited
leptons as

\begin{equation}
\Gamma(l^{\star}\rightarrow lV)=\frac{\alpha m^{\star3}}{4\Lambda^{2}}f_{V}^{2}(1-\frac{m_{V}^{2}}{m^{\star2}})^{2}(1+\frac{m_{V}^{2}}{2m^{\star2}}),
\end{equation}
where $f_{V}$ is the new electroweak coupling parameter corresponding
to the gauge boson $V$ and $f_{\gamma}=(f-f^{\prime})/2,$ $f_{Z}=(fcot\theta_{W}+f^{\prime}tan\theta_{W})/2,$
$f_{W}=f/\sqrt{2}sin\theta_{W}$, where $\theta_{W}$ is the weak
mixing angle, and $m_{V}$ is the mass of the gauge boson. The total
decay widths of the excited neutrino for the scale of $\varLambda=m_{\nu^{\star}}$
is given in Fig. 2.

\begin{figure}
\begin{centering}
\includegraphics[scale=0.6]{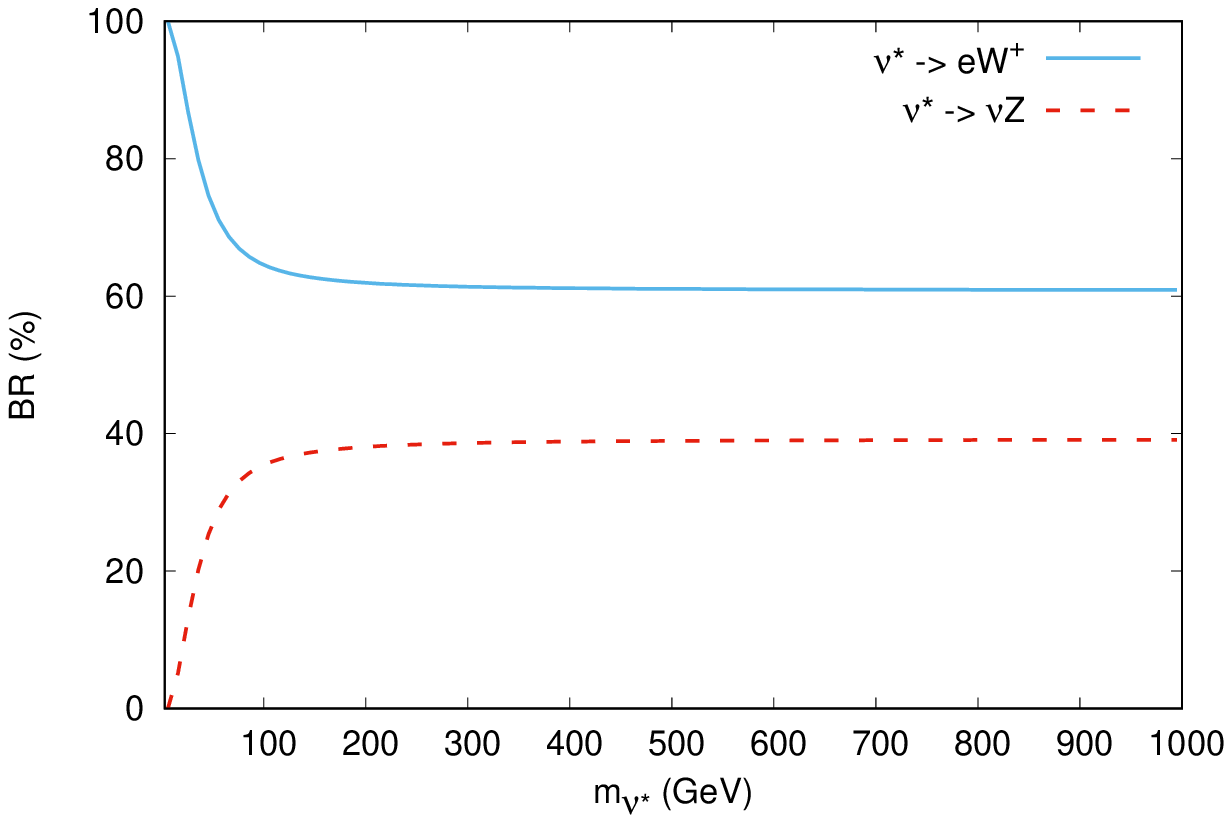}\includegraphics[scale=0.6]{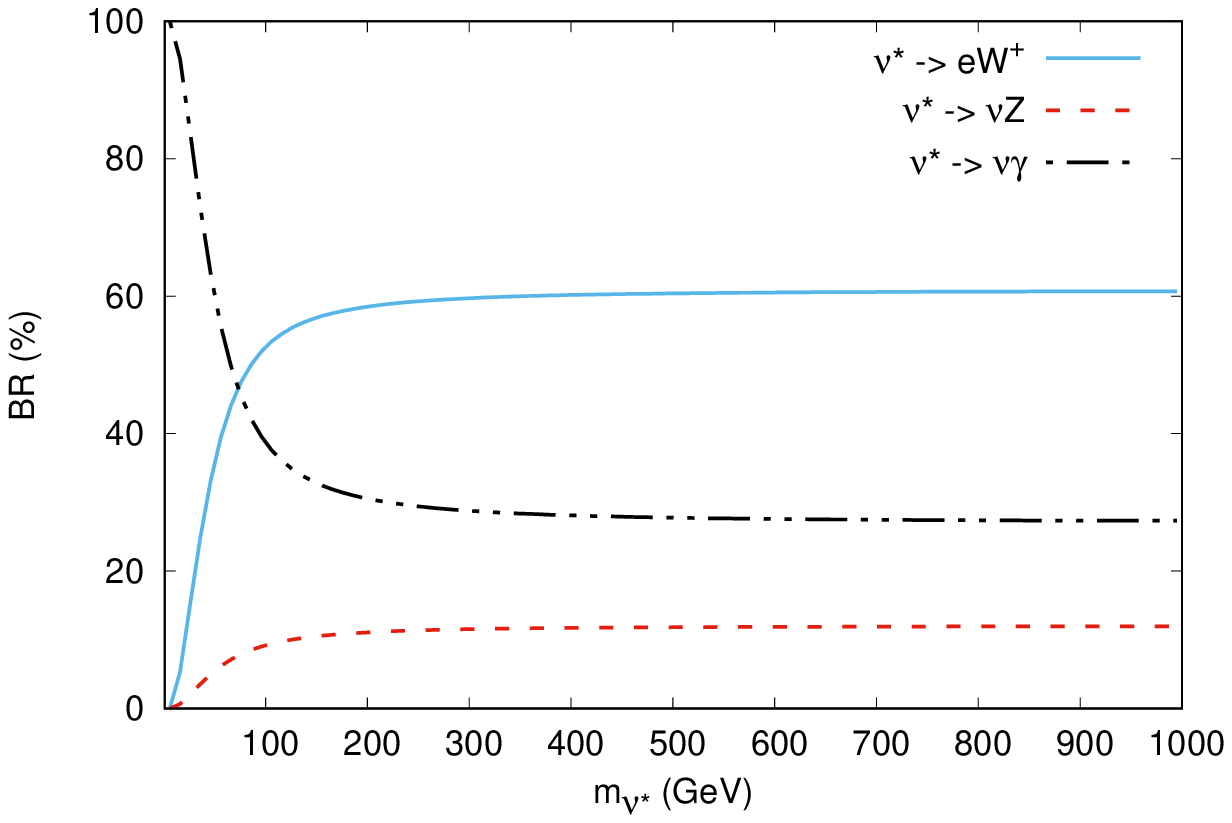}
\par\end{centering}
\caption{The branching ratios (\%) depending on the mass of the excited neutrino
for $f=f^{\prime}=1$ (left) and $f=-f^{\prime}=1$ (right).}
\end{figure}

\begin{figure}[H]
\begin{centering}
\includegraphics[scale=0.7]{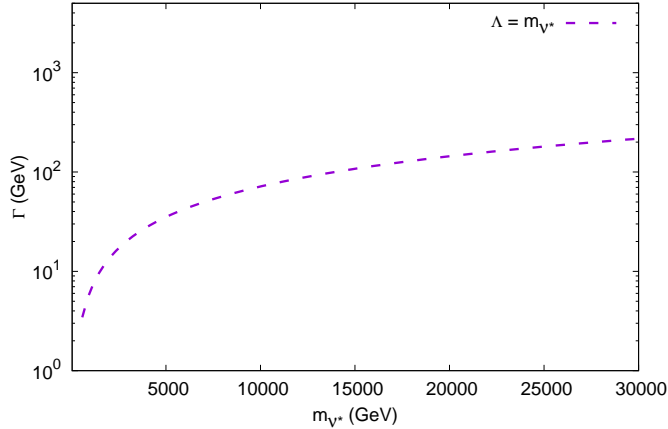}
\par\end{centering}
\caption{The total decay widths of the excited neutrino for the scale of $\varLambda=m_{\nu^{\star}}$
with the coupling of $f=f^{\prime}=1$. }
\end{figure}

\section{SIGNAL AND BACKGROUND ANALYSIS}

We analyze the potentials of the future $ep$ collider machines to
search for the excited neutrinos via the single production reaction
$ep\rightarrow\nu^{\star}X$ with subsequent decay of the excited
neutrino into an electron and $W^{+}$ boson. So, we deal with the
process $ep\rightarrow W^{+}eX$ and subprocesses $eq(\overline{q})\rightarrow W^{+}eq(\overline{q})$.
The signal and background analysis were done at the parton level by
using the high energy simulation program of CALCHEP (ver. 3.6.25)
\cite{calchep}. In our calculations we have used the parton distribution
functions library of CTEQ6L \cite{cteq}. 

For a comparison of different FCC-based $ep$ colliders, the signal
cross sections for excited neutrino production are presented in Fig.
3, assuming the coupling parameter of $f=f^{\prime}=1$. 

\begin{figure}[H]
\begin{centering}
\includegraphics[scale=0.85]{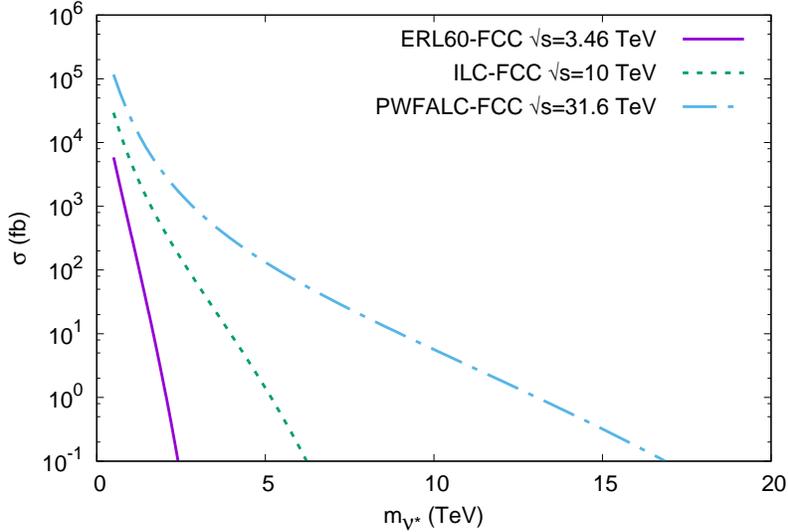}
\par\end{centering}
\caption{The total cross sections as a function of the excited neutrino mass
with the coupling of $f=f^{\prime}=1$, and the energy scale of $\varLambda=m_{\nu^{\star}}$
at the $ep$ colliders with various center-of-mass energies. }
\end{figure}

\subsection{ERL60$\otimes$FCC Collider}

The machine of ERL60$\otimes$FCC is a FCC-based future $ep$ collider
with the center-of-mass energy of $3.46$ TeV. Keeping in mind that
the lower bound on the mass of the excited neutrino is $1.6$ TeV
($m_{\nu^{\star}}>1.6$ $TeV$), we have explored the mass limits
for discovery of the excited neutrinos in the range of $1.6$ and
$3.46$ TeV for the ERL60$\otimes$FCC collider. Firstly, we have
applied initial kinematical cuts on the final state particles (electron,
$W^{+}$ boson, and jets) to form signal and backgrounds as $p_{T}^{e,W,j}>20$
GeV, where the $p_{T}$ is the transverse momentum of the final state
detectable particles. The SM cross section after the application of
these cuts has been calculated as $\sigma_{B}=3.96$ pb. In order
to define the kinematical cuts best suited for discovery, we have
plotted the normalized transverse momentum and the normalized pseudorapidity
distributions of the final state particles. Fig. 4 shows the $p_{T}$
distributions of the final state $W^{+}$ bosons and the $\eta$ (pseudorapidity)
distributions of the final state electrons for the excited neutrino
masses of $1000$ and $2000$ GeV versus the backgrounds. The $p_{T}$
distributions of $W^{+}$ bosons are the same for the final state
electrons. As can be seen from the Fig. 4, the selection of the kinematical
cuts as $p_{T}^{W}>200$ GeV (same for the electron) and $-5<\eta^{e}<-1$,
drastically reduces the background while keeping the signal almost
unaffected. As for the kinematical cut of $\eta$ distributions (see
Fig. 5 (right)) of final state $W^{+}$ bosons, it was determined
as $-4.5<\eta^{W}<-2$. The Fig. 5 (left) shows the invariant mass
distributions of the $eW^{+}$ system after application of the all
kinematical cuts for discovery. It is clearly seen that the background
is suppressed.

\begin{figure}
\begin{centering}
\includegraphics[scale=0.45]{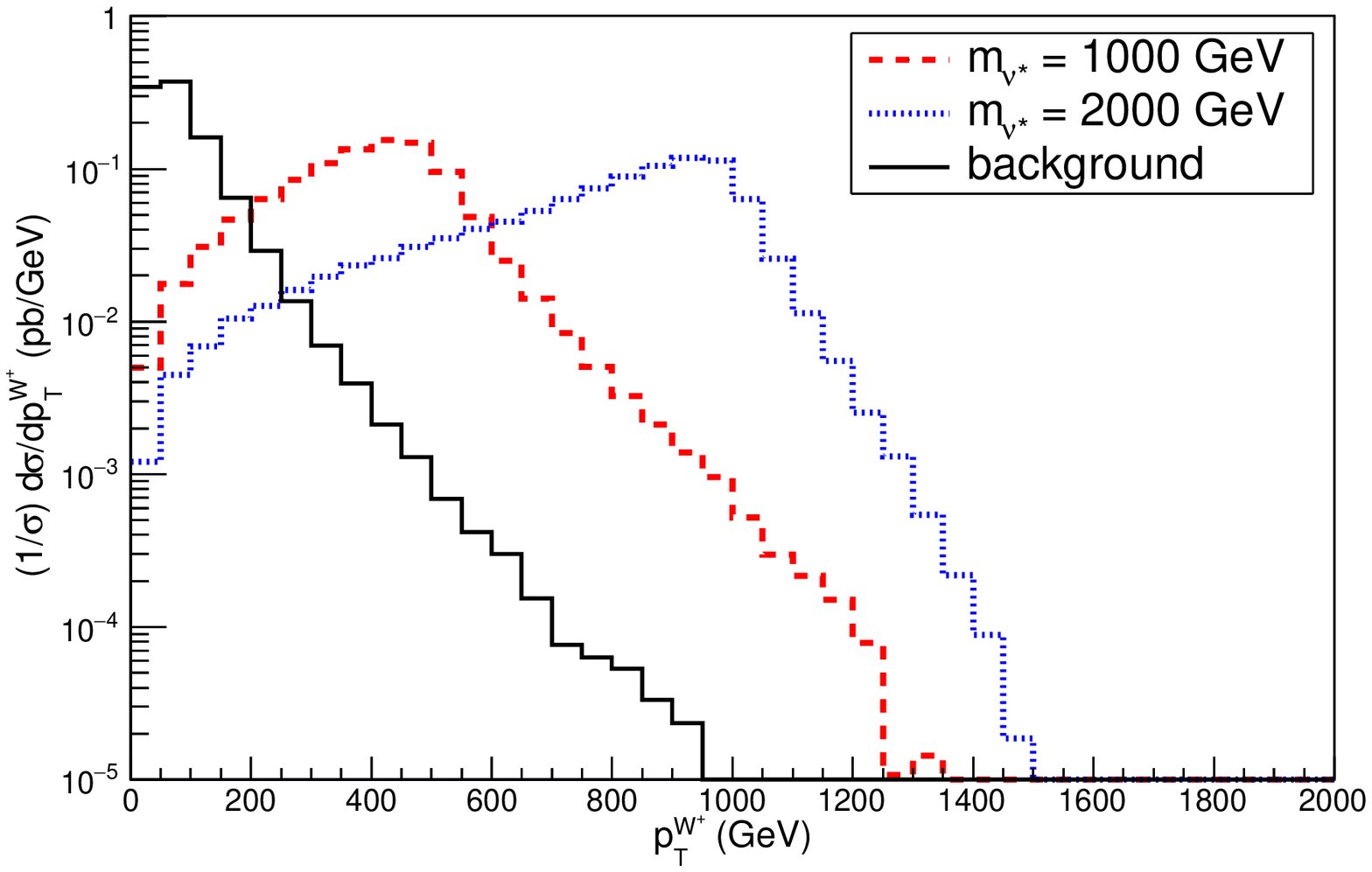}\includegraphics[scale=0.45]{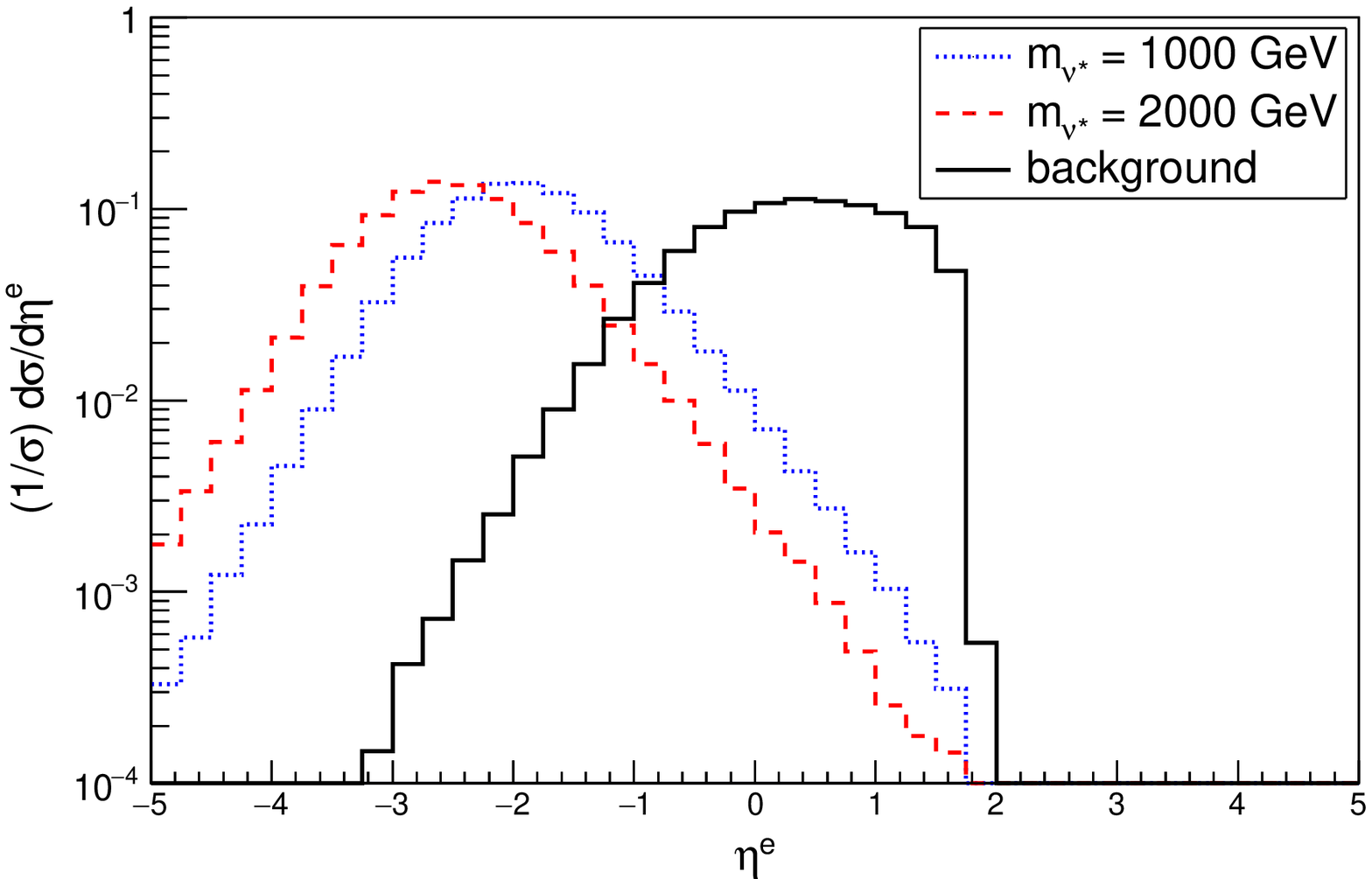}
\par\end{centering}
\caption{The normalized transverse momentum distributions of the final state
$W^{+}$ bosons (left) and the normalized pseudorapidity distributions
of the final state electrons (right) with the coupling of $f=f^{\prime}=1$
and the energy scale of $\varLambda=m_{\nu^{\star}}$ at the ERL60$\otimes$FCC
collider.}

\end{figure}

A natural way of extracting the excited neutrino signal, and the same
time suppressing the SM background is to impose a cut on the $eW^{+}$
invariant mass in addition to kinematical cuts for discovery. Therefore,
we have specified the cuts of mass window as $m_{\nu^{\star}}-2\Gamma_{\nu^{\star}}<m_{eW}<m_{\nu^{\star}}+2\Gamma_{\nu^{\star}}$. 

\begin{figure}
\begin{centering}
\includegraphics[scale=0.45]{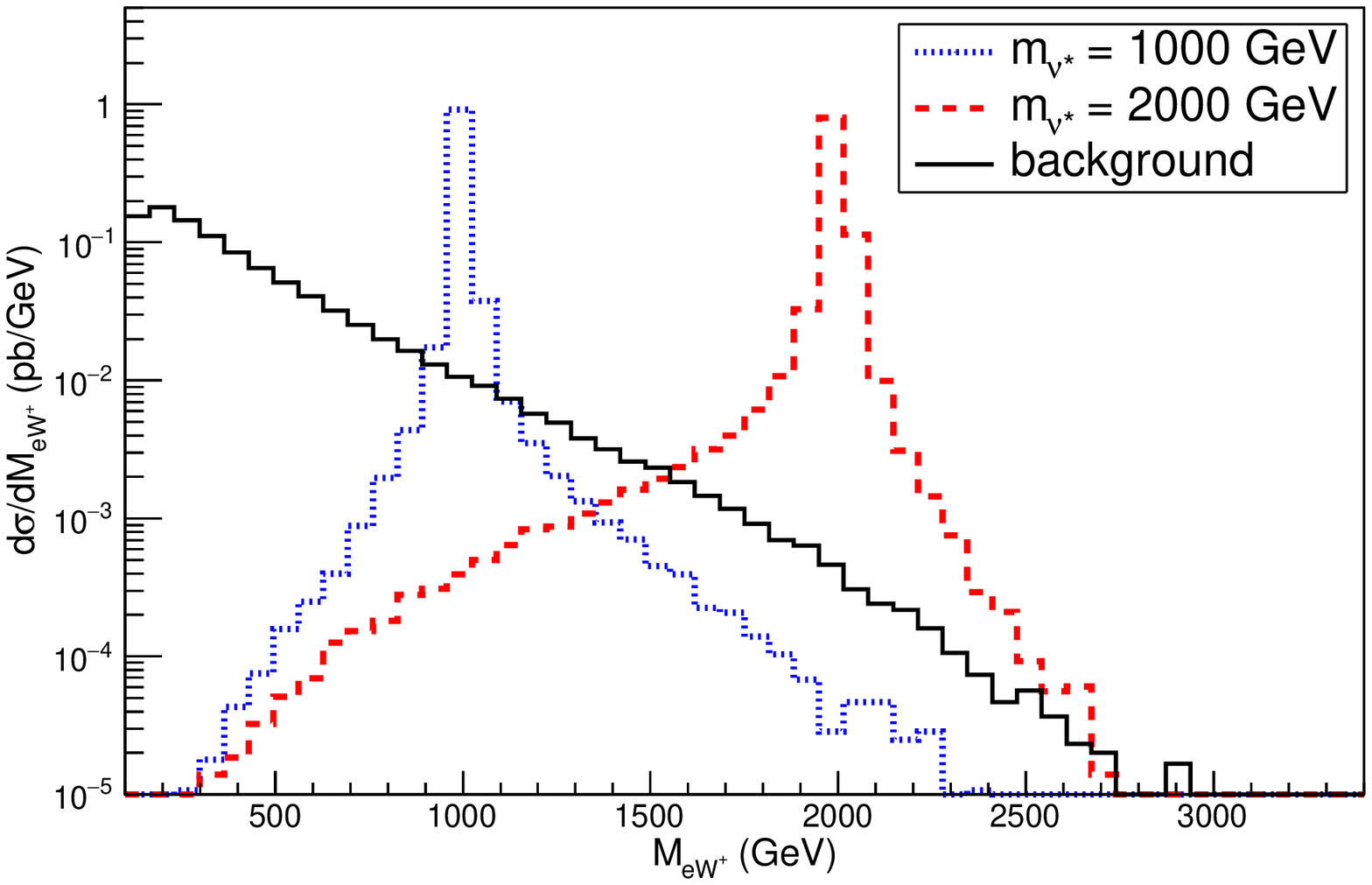}\includegraphics[scale=0.45]{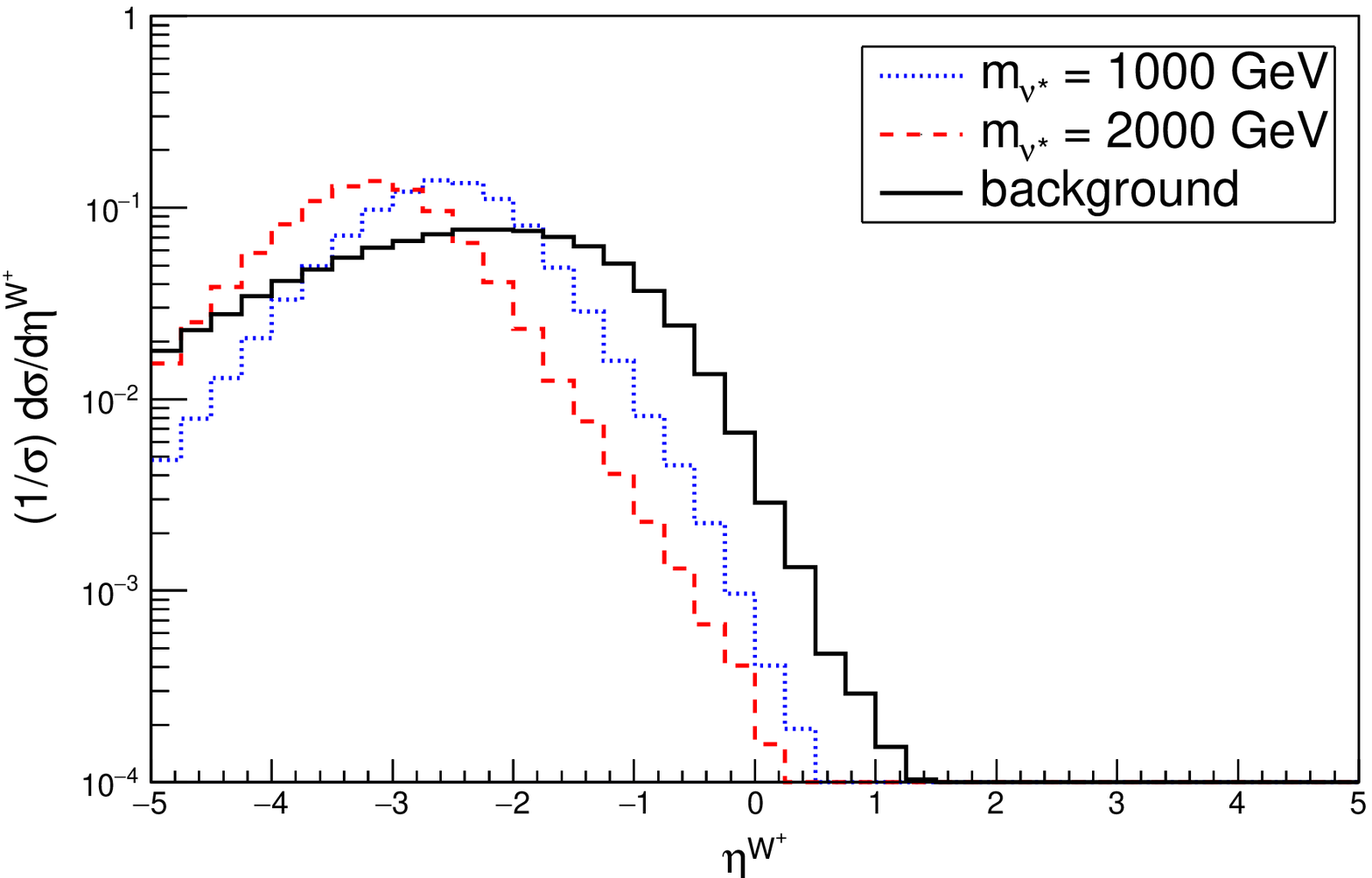}
\par\end{centering}
\caption{The invariant mass distributions of the excited neutrino signal and
the corresponding background (left), and the normalized pseudorapidity
distributions of the final state $W^{+}$ bosons (right), with the
energy scale of $\varLambda=m_{\nu^{\star}}$ and the coupling of
$f=f^{\prime}=1$ at the ERL60$\otimes$FCC collider. }

\end{figure}

By using the all kinematical cuts, we have calculated the statistical
significance (SS) values of the expected signal yield using the following
formula, 

\begin{equation}
SS=\frac{|\sigma_{S+B}-\sigma_{B}|}{\sqrt{\sigma_{B}}}\sqrt{L_{int}},
\end{equation}

where $L_{int}$ is the integrated luminosity of the collider. In
the Table 2, we have presented the signal (with the coupling of $f=f^{\prime}=1$
and the energy scale of $\varLambda=m_{\nu^{\star}}$ ) and the background
cross sections in $eW^{+}$ invariant mass bins since the signal is
concentrated in a small region proportional to the invariant mass
resolution. As can be understood from the Table II, the ERL60$\otimes$FCC
collider can discover the excited neutrino in $\nu^{\star}\rightarrow W^{+}e$
decay mode for the coupling of $f=f^{\prime}=1$ up to the mass of
$2452$ GeV taking into account the discovery criteria ($SS\geq5$).

\begin{table}[H]
\caption{The statistical significance (SS) values and the cross sections of
the excited neutrino signal and relevant backgrounds at ERL60$\otimes$FCC
collider with $\sqrt{s}=3.46$ TeV and $L_{int}=100$ $fb^{-1}$ assuming
the energy scale of $\varLambda=m_{\nu^{\star}}$ and the coupling
of $f=f^{\prime}=1$.}
\centering{}$ $%
\begin{tabular}{|c|c|c|c|}
\hline 
Mass (GeV) & $\sigma_{S+B}$ (pb) & $\sigma_{B}$ (pb) & SS\tabularnewline
\hline 
\hline 
$1600$ & $7.21$ x $10^{-3}$ & $1.86$ x $10^{-4}$ & $162.9$\tabularnewline
\hline 
$1800$ & $2.47$ x $10^{-3}$ & $9.60$ x $10^{-5}$ & $76.5$\tabularnewline
\hline 
$2000$ & $7.65$ x $10^{-4}$ & $4.15$ x $10^{-5}$ & $35.5$\tabularnewline
\hline 
$2200$ & $2.09$ x $10^{-4}$ & $1.49$ x $10^{-5}$ & $15.9$\tabularnewline
\hline 
$2300$ & $1.03$ x $10^{-4}$ & $8.48$ x $10^{-6}$ & $10.2$\tabularnewline
\hline 
$2400$ & $4.82$ x $10^{-5}$ & $4.64$ x $10^{-6}$ & $6.4$\tabularnewline
\hline 
$2500$ & $2.14$ x $10^{-5}$ & $2.41$ x $10^{-6}$ & $3.8$\tabularnewline
\hline 
$2600$ & $8.85$ x $10^{-6}$ & $1.19$ x $10^{-6}$ & $2.2$\tabularnewline
\hline 
$2700$ & $3.32$ x $10^{-6}$ & $5.41$ x $10^{-7}$ & $1.1$\tabularnewline
\hline 
\end{tabular}
\end{table}

\subsection{ILC$\otimes$FCC Collider}

The ILC$\otimes$FCC collider with the center-of-mass energy of $10$
TeV can search the excited neutrino in a wider mass range compared
to the ERL60$\otimes$FCC collider. We have explored the mass limits
for discovery of the excited neutrinos in the range of $1.6$ and
$10$ TeV. In order to perceive the excited neutrino signals from
the background we have put the same initial kinematical cuts, namely
$p_{T}^{e,W,j}>20$ GeV, with the ERL60$\otimes$FCC collider. The
SM background cross section for the ILC$\otimes$FCC collider is found
to be $\sigma_{B}=15.74$ pb after the application of these cuts.
The normalized transverse momentum distributions of the final state
electrons and the pseudorapidity distributions of the final state
$W^{+}$ bosons are presented in Fig. 6. For these distributions,
we have determined the kinematical cuts for discovery as $p_{T}^{e}>200$
GeV and $-3.4<\eta^{W}<0.4$. Transverse momentum distributions and
its kinematical cuts of the final state electrons and $W^{+}$ bosons
are the same. Fig. 7 shows the normalized pseudorapidity distributions
of the final state electrons, and the invariant mass distributions
of the $eW^{+}$ system after application of the all kinematical cuts
for discovery. The kinematical discovery cut of this distributions
was determined as $-5<\eta^{e}<1$.

\begin{figure}
\begin{centering}
\includegraphics[scale=0.45]{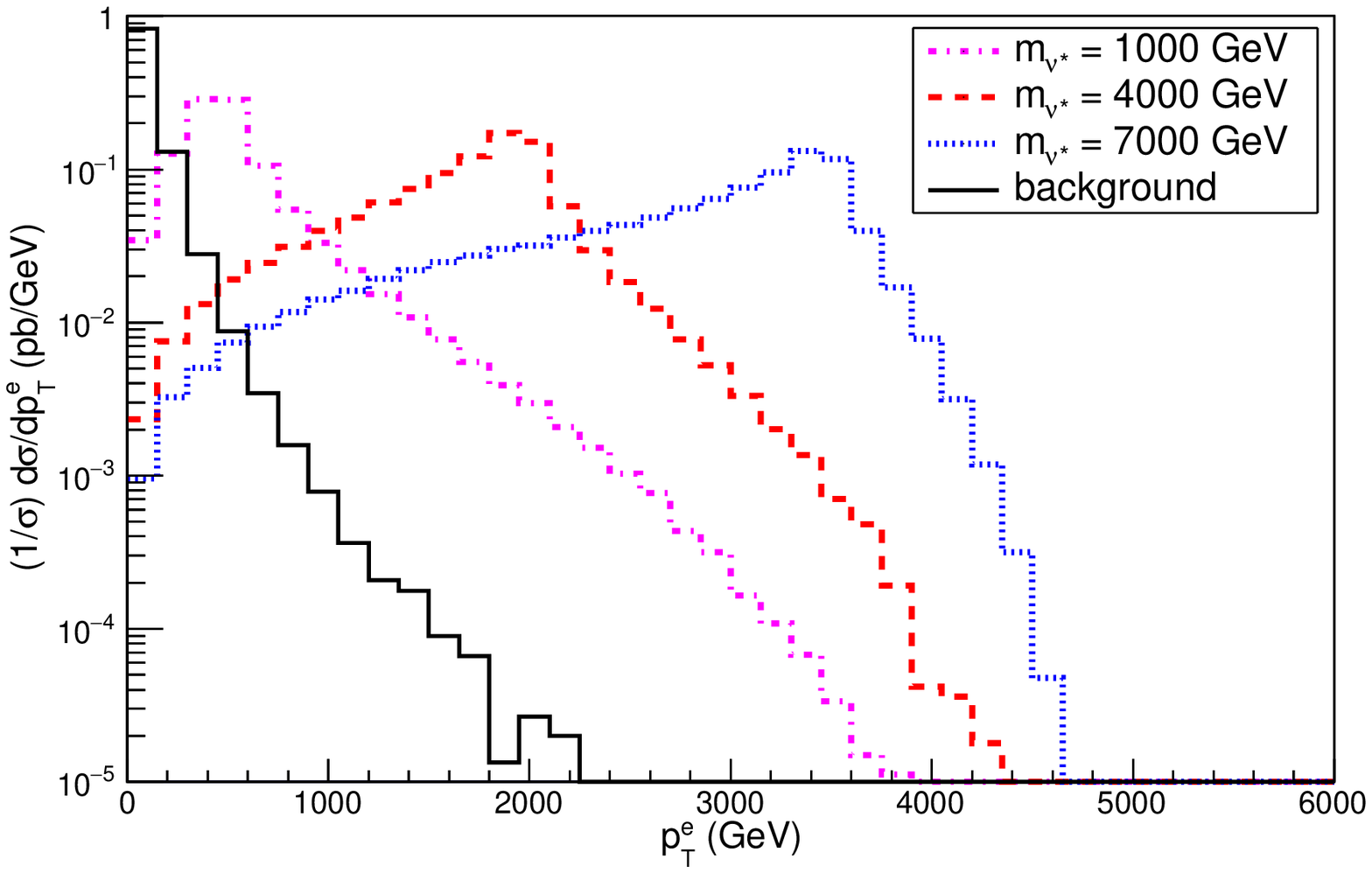}\includegraphics[scale=0.45]{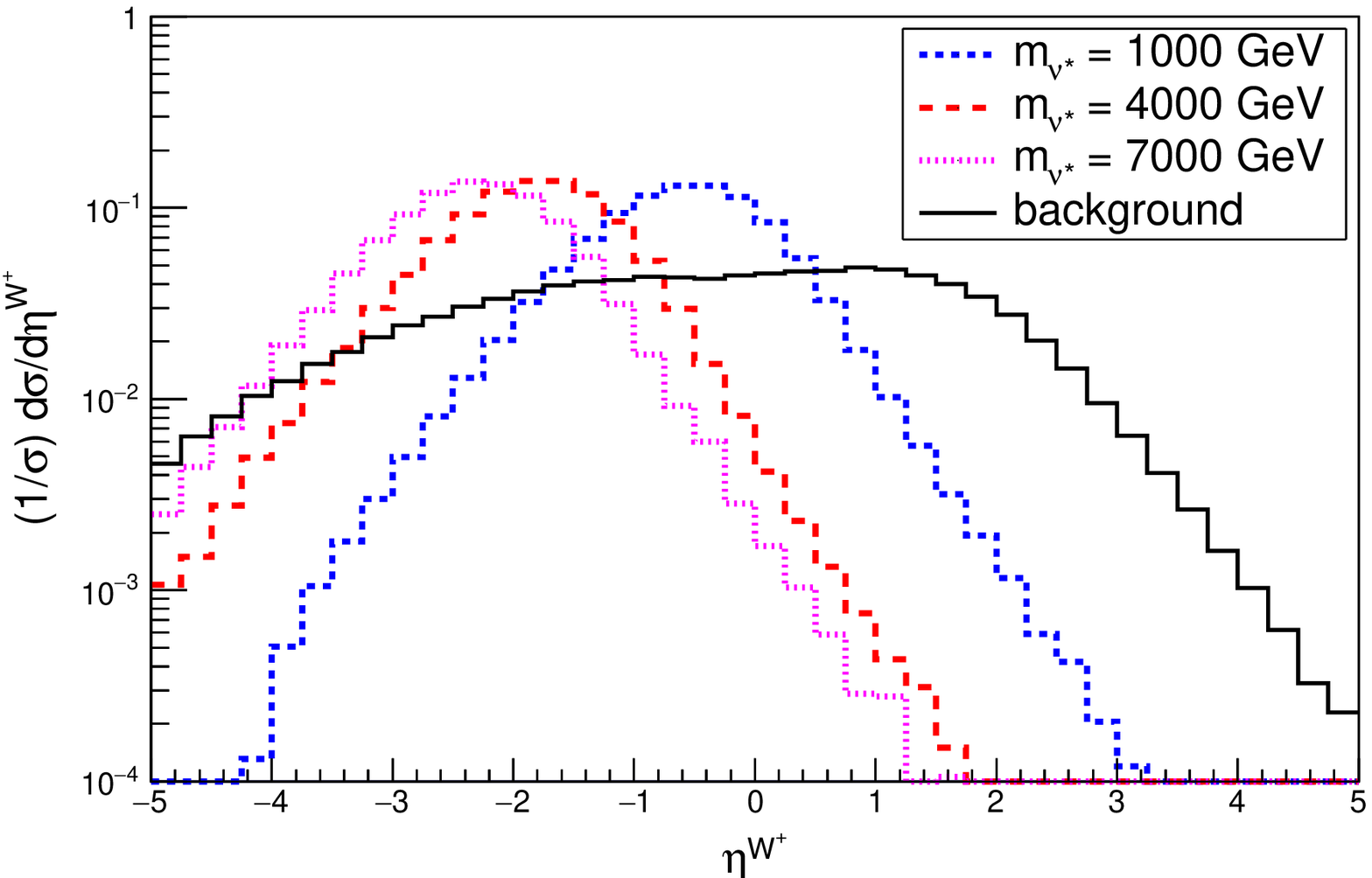}
\par\end{centering}
\caption{The normalized transverse momentum distributions of the final state
electrons (left) and the normalized pseudorapidity distributions of
the final state $W^{+}$ bosons (right) with the coupling of $f=f^{\prime}=1$
and the energy scale of $\varLambda=m_{\nu^{\star}}$ at the ILC$\otimes$FCC
collider.}
\end{figure}

Table 3 presents the signal and background cross sections in $e$$W^{+}$
invariant mass bins satisfying the condition of $m_{\nu^{\star}}-2\Gamma_{\nu^{\star}}<m_{eW}<m_{\nu^{\star}}+2\Gamma_{\nu^{\star}}$. 

\begin{figure}
\begin{centering}
\includegraphics[scale=0.45]{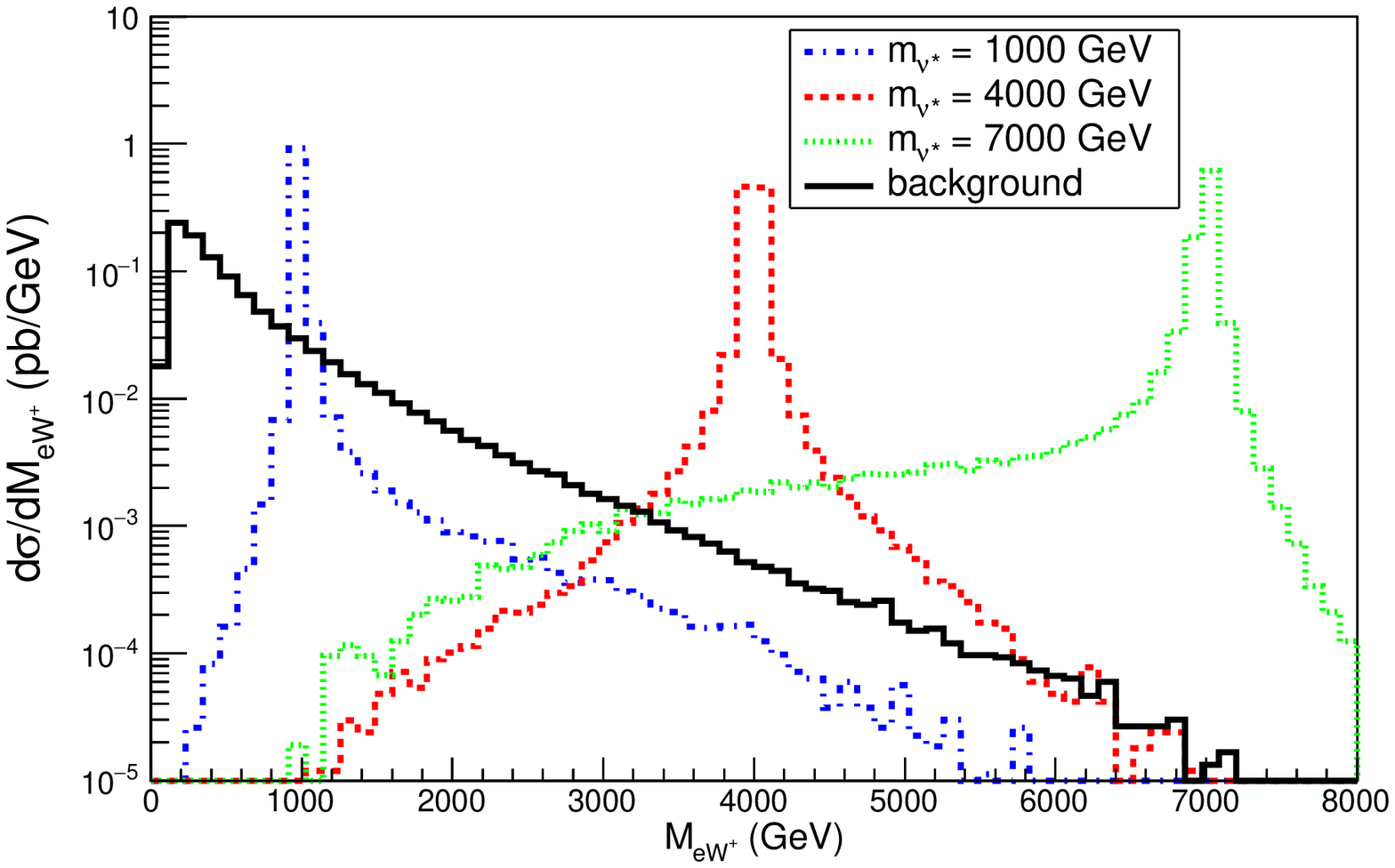}\includegraphics[scale=0.45]{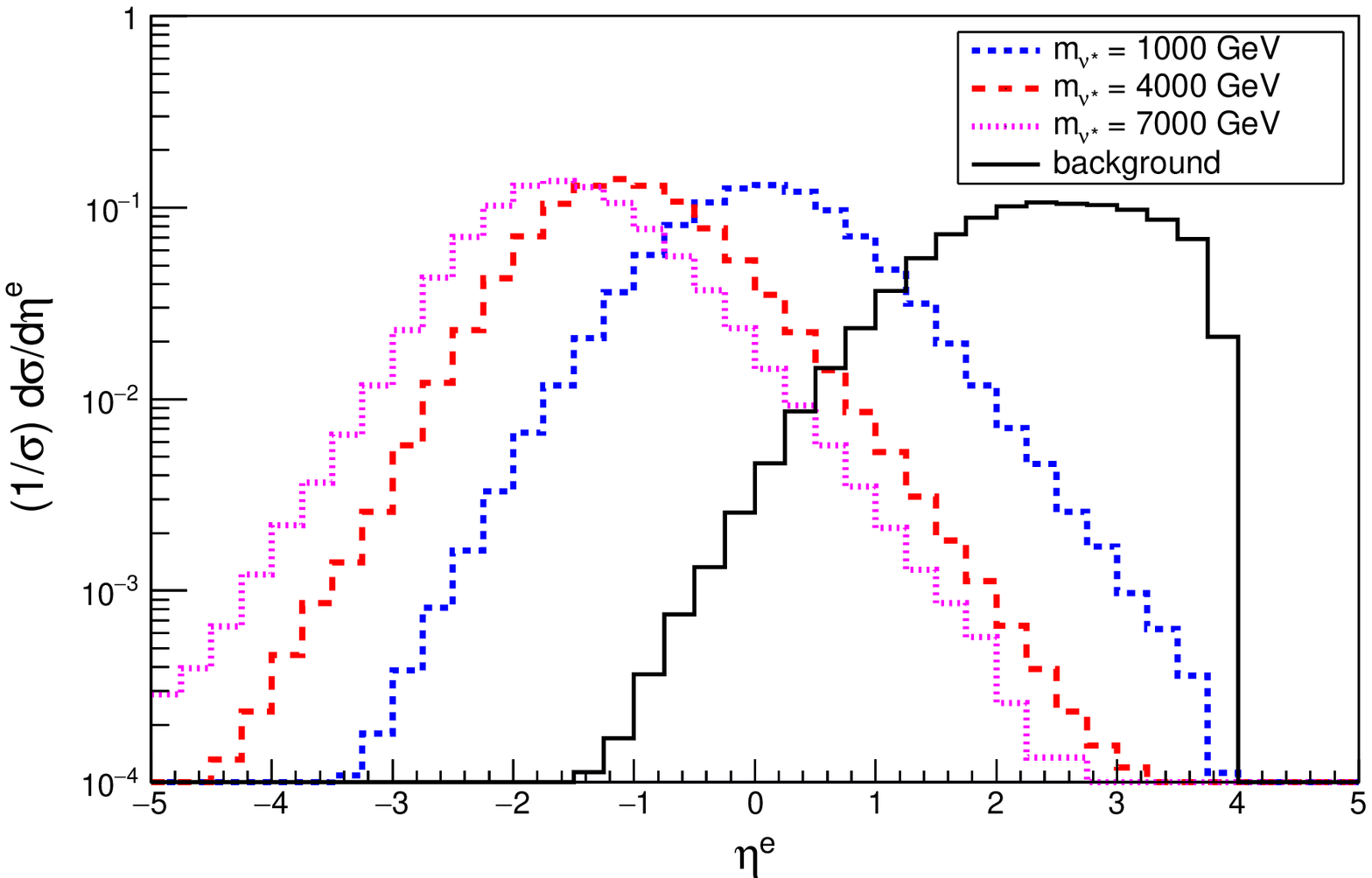}
\par\end{centering}
\caption{The invariant mass distributions of the excited neutrino signal and
the corresponding background (left), and the normalized pseudorapidity
distributions of the final state electrons (right), with the energy
scale of $\varLambda=m_{\nu^{\star}}$ and the coupling of $f=f^{\prime}=1$
at the ILC$\otimes$FCC collider. }

\end{figure}

When we look at the calculated SS values for $SS\geq5$ criteria in
Table III, for the energy scale of $\varLambda=m_{\nu^{\star}}$,
the ILC$\otimes$FCC collider can probe the excited neutrino (assuming
the coupling of $f=f^{\prime}=1$) up to the masses of $5635$ and
$6460$ GeV for the integrated luminosities of $L_{int}=10$ $fb^{-1}$
and $L_{int}=100$ $fb^{-1}$ , respectively.

\begin{table}[H]
\caption{The statistical significance (SS) values and the cross sections of
the excited neutrino signal and relevant background at the ILC$\otimes$FCC
collider with $\sqrt{s}=10$ TeV assuming the coupling of $f=f^{\prime}=1$
and the energy scale of $\varLambda=m_{\nu^{\star}}$.}
\centering{}%
\begin{tabular}{|c|c|c|c|c|}
\hline 
\multirow{2}{*}{Mass (GeV)} & \multirow{2}{*}{$\sigma_{B}$ (pb)} & \multirow{2}{*}{$\sigma_{S+B}$ (pb)} & $L_{int}=10$ $fb^{-1}$  & $L_{int}=100$ $fb^{-1}$ \tabularnewline
\cline{4-5} 
 &  &  & SS & SS \tabularnewline
\hline 
$2000$ & $1.81$ x $10^{-3}$ & $1.47$ x $10^{-1}$ & $342.1$ & $1081.9$\tabularnewline
\hline 
$2500$ & $1.24$ x $10^{-3}$ & $5.85$ x $10^{-2}$ & $162.9$ & $515.1$\tabularnewline
\hline 
$3000$ & $7.37$ x $10^{-4}$ & $2.43$ x $10^{-2}$ & $86.8$ & $274.7$\tabularnewline
\hline 
$3500$ & $3.94$ x $10^{-4}$ & $1.03$ x $10^{-2}$ & $49.6$ & $157$\tabularnewline
\hline 
$4000$ & $1.85$ x $10^{-4}$ & $4.28$ x $10^{-3}$ & $30$ & $95.1$\tabularnewline
\hline 
$4500$ & $8.27$ x $10^{-5}$ & $1.74$ x $10^{-3}$ & $18.1$ & $57.4$\tabularnewline
\hline 
$5000$ & $3.58$ x $10^{-5}$ & $6.69$ x $10^{-4}$ & $10.5$ & $33.4$\tabularnewline
\hline 
$5500$ & $1.47$ x $10^{-5}$ & $2.43$ x $10^{-4}$ & $5.9$ & $18.8$\tabularnewline
\hline 
$6000$ & $6.07$ x $10^{-6}$ & $8.15$ x $10^{-5}$ & $3$ & $9.6$\tabularnewline
\hline 
$6500$ & $2.26$ x $10^{-6}$ & $2.49$ x $10^{-5}$ & $1.5$ & $4.7$\tabularnewline
\hline 
$7000$ & $7.83$ x $10^{-7}$ & $6.69$ x $10^{-6}$ & $0.6$ & $2.1$\tabularnewline
\hline 
$7500$ & $2.37$ x $10^{-7}$ & $1.49$ x $10^{-6}$ & $0.2$ & $0.8$\tabularnewline
\hline 
\end{tabular}
\end{table}

\subsection{PWFA-LC$\otimes$FCC Collider}

If the excited neutrinos had not been observed at the ERL60$\otimes$FCC
and the ILC$\otimes$FCC colliders, they would have been explored
up to the mass of $31.6$ TeV at the PWFA-LC$\otimes$FCC collider
that has the widest research potential. We have explored the mass
limits for discovery of the excited neutrinos in a broad mass spectrum
from $1.6$ to $31.6$ TeV. The SM background cross section is found
to be $\sigma_{B}=58.15$ pb after application of the same initial
kinematical cuts. Fig. 8 shows the $p_{T}$ distributions of the final
state $W^{+}$ bosons and the $\eta$ distributions of the final state
electrons for the excited neutrino masses of $5000$, $10000$, $15000$,
and $20000$ GeV versus the backgrounds. $p_{T}$ distributions of
the $W^{+}$ bosons are the same for the final state electrons. It
is seen from the Fig. 8 that the selection of the kinematical cuts
as $p_{T}^{W}>400$ GeV (same for the electron) and $-5<\eta^{e}<2.5$,
essentially suppress the background, whereas the signal remains almost
unchanged. The normalized pseudorapidity distributions of the $W^{+}$
bosons, and the invariant mass distributions of the $eW^{+}$ system
obtained after application of the all discovery cuts are given in
Fig. 9. According to this Figure, the discovery cut of the normalized
pseudorapidity distributions of the final state $W^{+}$ bosons was
determined as $-2.5<\eta^{W}<1$. In addition to these cuts, we have
also applied the cuts to the $eW^{+}$ invariant masses using the
$m_{\nu^{\star}}-2\Gamma_{\nu^{\star}}<m_{eW}<m_{\nu^{\star}}+2\Gamma_{\nu^{\star}}$. 

\begin{figure}
\begin{centering}
\includegraphics[scale=0.45]{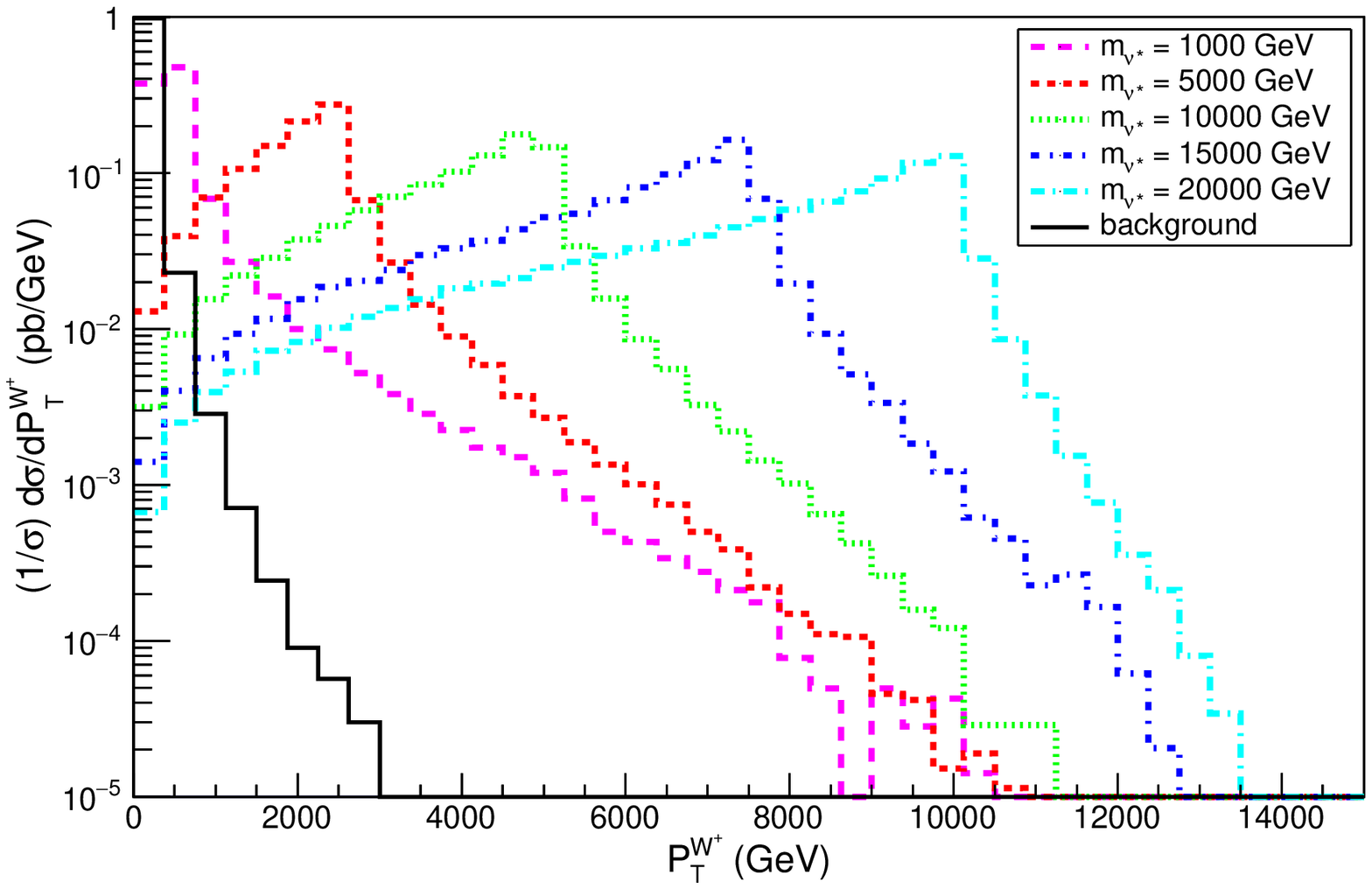}\includegraphics[scale=0.45]{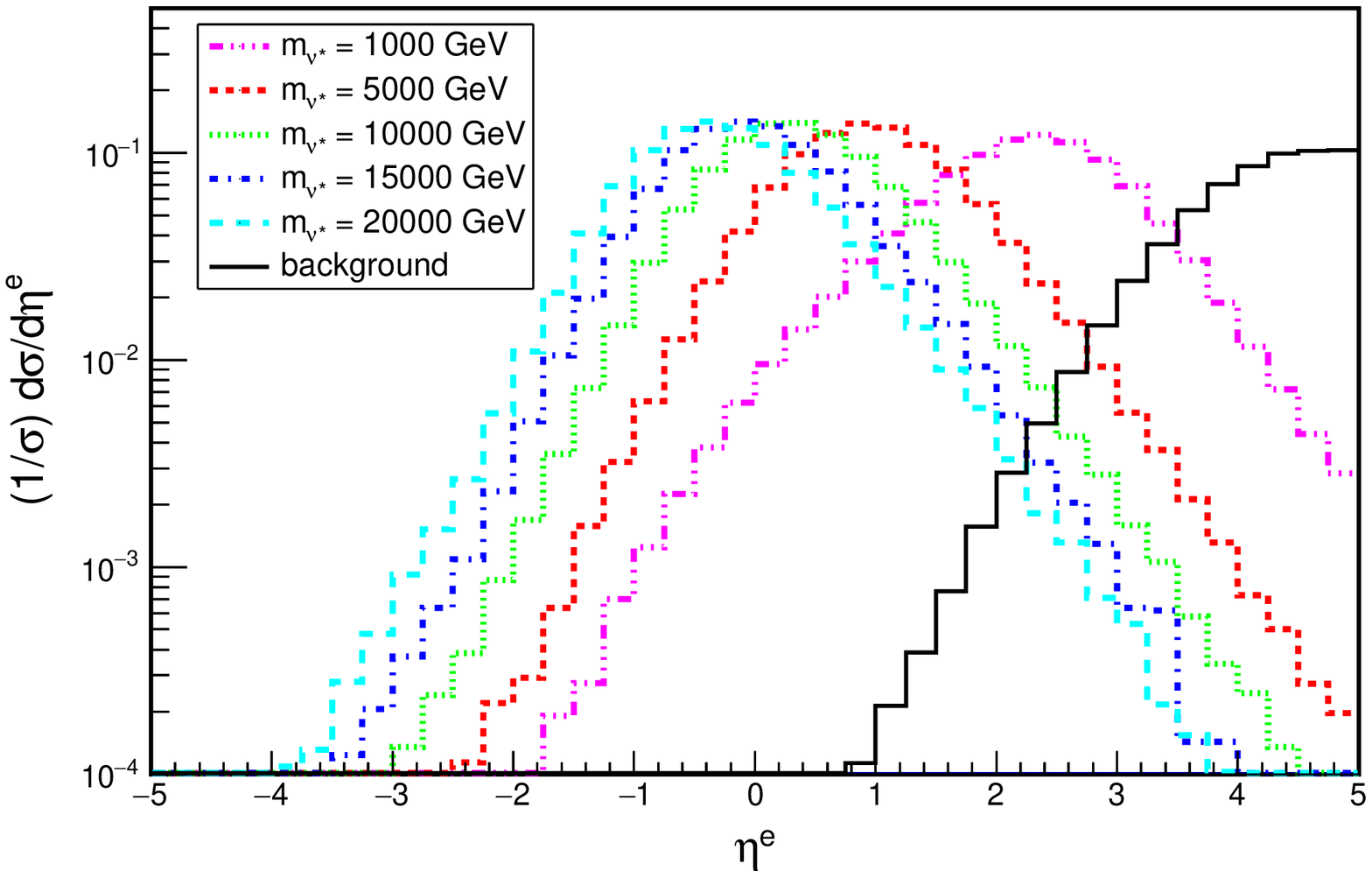}
\par\end{centering}
\caption{The normalized transverse momentum distributions of the final state
$W^{+}$ bosons (left) and the normalized pseudorapidity distributions
of the final state electrons (right) with the coupling of $f=f^{\prime}=1$
and the energy scale of $\varLambda=m_{\nu^{\star}}$ at the PWFA-LC$\otimes$FCC
collider.}
\end{figure}

\begin{figure}
\begin{centering}
\includegraphics[scale=0.45]{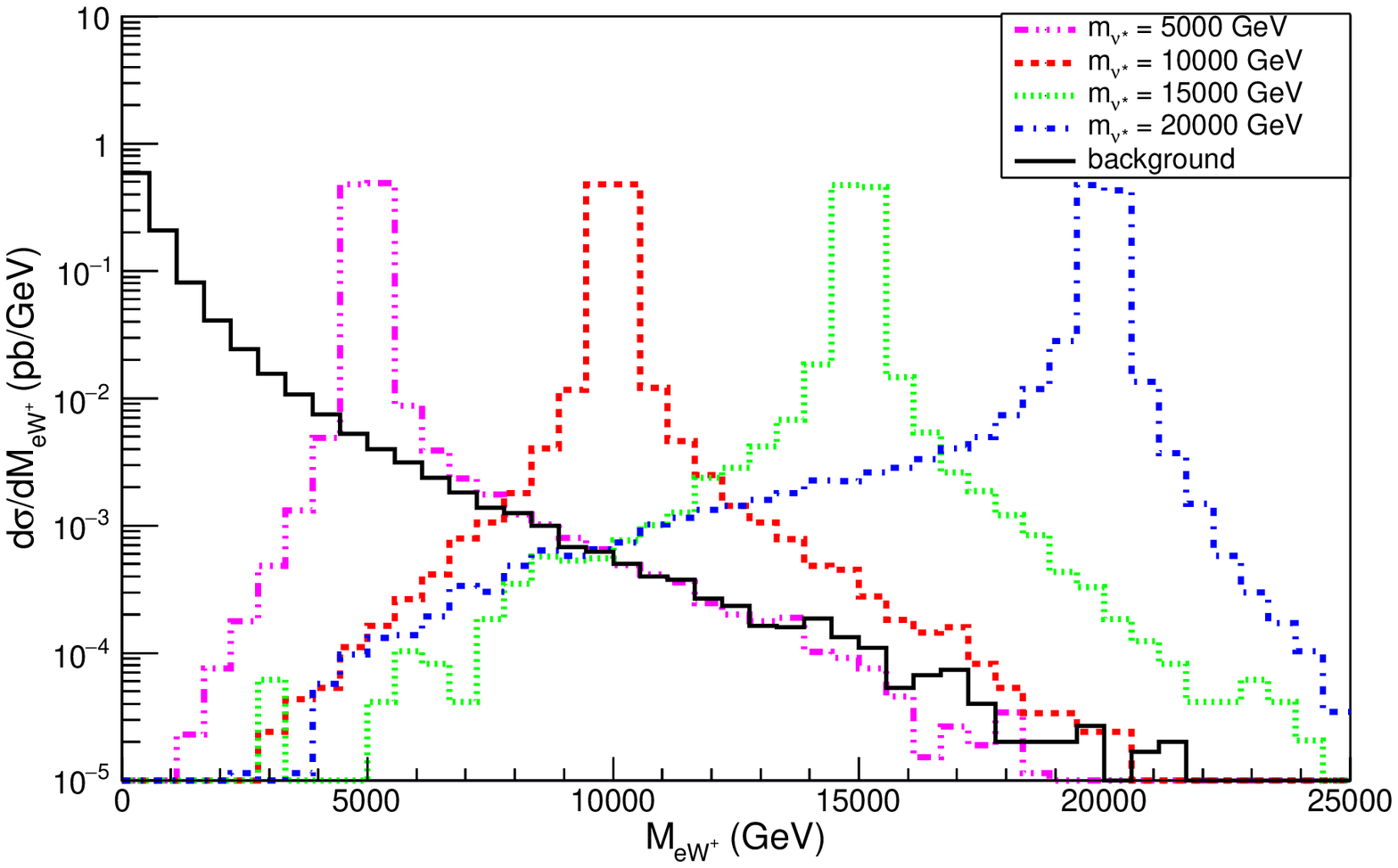}\includegraphics[scale=0.45]{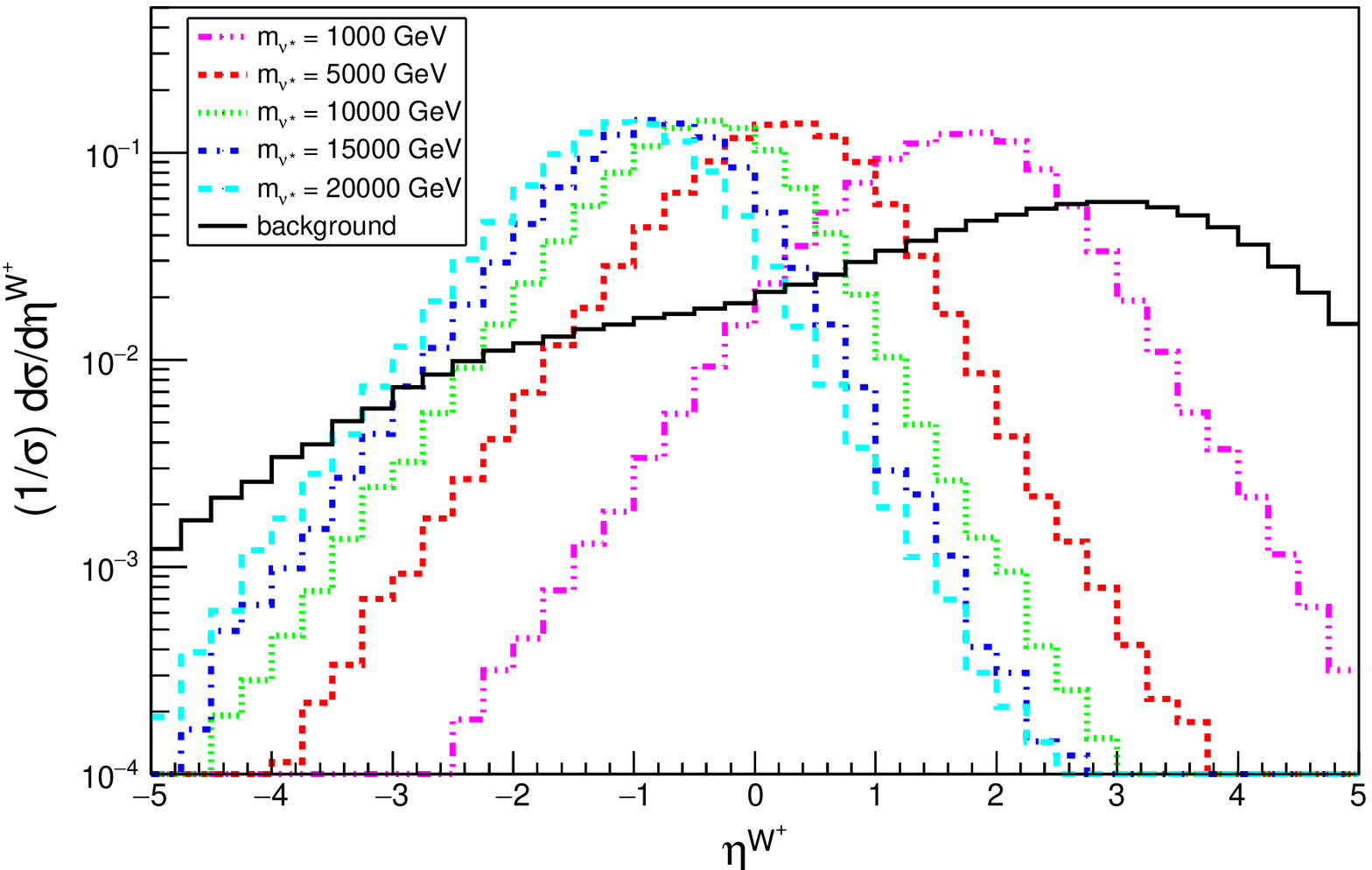}
\par\end{centering}
\caption{The invariant mass distributions of the excited neutrino signal and
the corresponding background (left), and the normalized pseudorapidity
distributions of the final state $W^{+}$ bosons (right), with the
energy scale of $\varLambda=m_{\nu^{\star}}$ and the coupling of
$f=f^{\prime}=1$ at the PWFA-LC$\otimes$FCC collider. }

\end{figure}

The signal and the background cross sections for PWFA-LC$\otimes$FCC
collider with the coupling of $f=f^{\prime}=1$ and the energy scale
of $\varLambda=m_{\nu^{\star}}$ are presented in Table 4 for two
integrated luminosity values, namely $L_{int}=1$ $fb^{-1}$ and $L_{int}=10$
$fb^{-1}$. For the energy scale of $\varLambda=m_{\nu^{\star}}$,
the PWFA-LC$\otimes$FCC collider can probe the excited neutrino up
to the masses of $10200$ and $13960$ GeV for the integrated luminosities
of $L_{int}=1$ $fb^{-1}$ and $L_{int}=10$ $fb^{-1}$ , respectively,
as can be understood from the Table 4.

\begin{table}[H]
\caption{The statistical significance (SS) values and the cross sections of
the excited neutrino signal and relevant background at the PWFA-LC$\otimes$FCC
collider with $\sqrt{s}=31.6$ TeV assuming the coupling of $f=f^{\prime}=1$
and the energy scale of $\varLambda=m_{\nu^{\star}}$. }
\centering{}%
\begin{tabular}{|c|c|c|c|c|}
\hline 
\multirow{2}{*}{Mass (GeV)} & \multirow{2}{*}{$\sigma_{B}$ (pb)} & \multirow{2}{*}{$\sigma_{S+B}$ (pb)} & $L_{int}=1$ $fb^{-1}$  & $L_{int}=10$ $fb^{-1}$ \tabularnewline
\cline{4-5} 
 &  &  & SS & SS \tabularnewline
\hline 
$2000$ & $1.16$ x $10^{-3}$ & $2.92$ x $10^{-1}$ & $270.5$ & $855.6$\tabularnewline
\hline 
$4000$ & $1.03$ x $10^{-3}$ & $9.00$ x $10^{-2}$ & $87.6$ & $277.2$\tabularnewline
\hline 
$6000$ & $6.30$ x $10^{-4}$ & $2.42$ x $10^{-2}$ & $29.7$ & $93.9$\tabularnewline
\hline 
$8000$ & $3.59$ x $10^{-4}$ & $7.47$ x $10^{-3}$ & $11.8$ & $37.5$\tabularnewline
\hline 
$10000$ & $1.78$ x $10^{-4}$ & $2.45$ x $10^{-3}$ & $5.3$ & $17$\tabularnewline
\hline 
$12000$ & $6.55$ x $10^{-5}$ & $8.02$ x $10^{-4}$ & $2.8$ & $9.1$\tabularnewline
\hline 
$14000$ & $2.27$ x $10^{-5}$ & $2.57$ x $10^{-4}$ & $1.5$ & $4.9$\tabularnewline
\hline 
$16000$ & $7.71$ x $10^{-6}$ & $7.66$ x $10^{-5}$ & $0.7$ & $2.4$\tabularnewline
\hline 
$18000$ & $2.51$ x $10^{-6}$ & $2.06$ x $10^{-5}$ & $0.3$ & $1.1$\tabularnewline
\hline 
$20000$ & $7.42$ x $10^{-7}$ & $4.85$ x $10^{-6}$ & $0.1$ & $0.4$\tabularnewline
\hline 
\end{tabular}
\end{table}

\section{CONCLUSION}

This work has shown that the FCC-based $ep$ colliders have a great
potential for the excited neutrino searches. We give the realistic
estimates for the excited neutrino signal and the corresponding background
at three different colliders, namely the ERL60$\otimes$FCC ($\sqrt{s}=3.46$
TeV), the ILC$\otimes$FCC ($\sqrt{s}=10$ TeV), and the PWFA-LC$\otimes$FCC
($\sqrt{s}=31.6$ TeV). The simulations have been performed for the
energy scale of $\varLambda=m_{\nu^{\star}}$ and the coupling parameter
of $f=f^{\prime}=1$. The mass limits for exclusion, observation,
and discovery of the excited neutrinos at the three colliders are
given in Table V, for the different integrated luminosity values.
As a result, these three FCC-based $ep$ colliders offer the possibility
to probe the excited neutrino over a very large mass range.

\begin{table}[H]
\caption{The mass limits for the exclusion ($2\sigma$), the observation ($3\sigma$),
and the discovery ($5\sigma$) of the excited neutrinos at the different
$ep$ colliders assuming the coupling of $f=f^{\prime}=1$ and the
energy scale of $\varLambda=m_{\nu^{\star}}$. }
\centering{}%
\begin{tabular}{|c|c|c|c|c|}
\hline 
\multirow{1}{*}{Colliders} & \multirow{1}{*}{$L_{int}$($fb^{-1}$) } & $2\sigma$ (GeV) & $3\sigma$ (GeV) & $5\sigma$ (GeV)\tabularnewline
\hline 
ERL60$\otimes$FCC & $100$ & $2618$ & $2547$ & $2452$\tabularnewline
\hline 
\multirow{2}{*}{ILC$\otimes$FCC} & $10$ & $6300$ & $6000$ & $5635$\tabularnewline
\cline{2-5} 
 & $100$ & $7025$ & $6790$ & $6460$\tabularnewline
\hline 
\multirow{2}{*}{PWFA-LC$\otimes$FCC} & $1$ & $13050$ & $11850$ & $10200$\tabularnewline
\cline{2-5} 
 & $10$ & $16500$ & $15450$ & $13960$\tabularnewline
\hline 
\end{tabular}
\end{table}
\begin{acknowledgments}
I am grateful to A. Ozansoy and S. O. Kara for useful discussions
and model file supports. This work has been supported by the Scientific
and Technological Research Council of Turkey (TUBITAK) under the grant
no 114F337.
\end{acknowledgments}


\begin{thebibliography}{10}
\bibitem[1]{ATLAS collaboration} ATLAS Collaboration, ``Observation
of a new particle in the search for the Standard Model Higgs boson
with the ATLAS detector at the LHC'', Phys. Lett. B 716, 1 (2012).

\bibitem[2]{I.A. DSouza} I.A. D'Souza and C.S. Kalman, PREONS: Models
of leptons, quarks and gauge bosons as composite objects, World Scientific
Publishing, 1992.

\bibitem[3]{J.H. K=0000FChn} J.H. Kühn, H.D. Tholl and P.M. Zerwas,
``Signals of excited quarks and leptons'', Phys. Lett. B, 158, 3
(1985).

\bibitem[4]{U.Baur} U. Baur, M. Spira and P.M. Zerwas, ``Excited-quark
and -lepton production at hadron colliders'', Phys. Rev. D 42, 815
(1990).

\bibitem[5]{Y. Tosa} Y. Tosa and R.E. Marshak, ``Exotic fermions'',
Phys. Rev. D 32, 774 (1985).

\bibitem[6]{LEP} L3 Collaboration, ``Search for excited leptons
at LEP'', Phys. Lett. B 568, 1 (2003).

\bibitem[7]{HERA} H1 Collaboration, ``Search for excited electrons
in ep collisions at HERA'', Phys. Lett. B 666, 2 (2008).

\bibitem[8]{Tevatron} D0 Collaboration, ``Search for excited electrons
in $p\bar{p}$ collision at $\sqrt{s}$$=$1.96 TeV'', Phys. Rev.
D 77, 091102 (2008).

\bibitem[9]{CMS} CMS Collaboration, ``Search for excited leptons
in proton-proton collisions at $\sqrt{s}=8$ TeV'', JHEP 2016, 125
(2016).

\bibitem[10]{ATLAS} ATLAS Collaboration, ``Search for excited electrons
and muons $\sqrt{s}=8$ TeV proton-proton collisions with the ATLAS
detector'', New J. Phys. 15, 093011 (2013).

\bibitem[11]{O. Cakir single production} O. Cakir, A. Yilmaz, S.
Sultansoy, ``Single production of excited electrons at future $e^{-}e^{+}$,
ep and pp colliders'', Phys. Rev. D 70, 075011 (2004).

\bibitem[12]{A.Ozansoy search} A. Ozansoy and A.A. Billur, ``Search
for excited electrons through $\gamma\gamma$ scattering'', Phys.
Rev. D 86, 055008 (2012).

\bibitem[13]{A.Caliskan} A. Caliskan, S.O. Kara, A. Ozansoy, ``Excited
muon searches at the FCC based muon-hadron colliders'', Adv. High
Energy Phys. 2017, 1540243 (2017).

\bibitem[14]{single production} O. Cakir, I.T. Cakir, Z. Kirca, ``Single
production of excited neutrinos at future $e^{+}e^{-}$, ep and pp
colliders'', Phys. Rev. D 70, 075017 (2004).

\bibitem[15]{3/2 neutrino} O. Cakir, A. Ozansoy, ``Single production
of excited spin-3/2 neutrinos at linear colliders'', Phys. Rev. D
79, 055001 (2009).

\bibitem[16]{k=0000F6ksal} M. Köksal, ``Analysis of excited neutrinos
at the CLIC'', Int. J. Mod. Phys. A29, 1450138 (2014).

\bibitem[17]{V.Ar=000131} A. Ozansoy, V. Ari, V. Cetinkaya, ``Search
for spin-3/2 neutrinos at LHeC'', Adv. High Energy Phys. 2016, 1739027
(2016). 

\bibitem[18]{particle sata group} C. Patrignani et al. (Particle
Data Group), ``Review of particle physics'', Chin. Phys. C 40, 100001
(2016).

\bibitem[19]{FCC web} FCC Project Web Page: https://fcc.web.cern.ch.

\bibitem[20]{TLEP} M. Bicer et al. (TLEP Working Group), ``First
look at the physics case of TLEP'', JHEP 1401, 164 (2014).

\bibitem[21]{Sultansoy rome} S. Sultansoy et al., ``FCC based lepton-hadron
and photon-hadron colliders: luminosity and physics'', Second Annual
Meeting of the Future Collider Study (FCC Week 2016), Rome, Italy
(2016).

\bibitem[22]{LHeC web} LHeC Project Web Page: http://lhec.web.cern.ch.

\bibitem[23]{ILC} C. Adolphsen et al., ``The International Linear
Collider Technical Design Report - Volume 3.II'', arXiv:1308.0494
{[}physics.acc-ph{]}.

\bibitem[24]{PWLC} J. P. Delahaye et al., ``A beam driven plasma-wakefield
linear collider from Higgs factory to multi - TeV'', in proceeding
of the Fifth International Particle Accelerator Conference, Dresden,
Germany, 3791 (2014).

\bibitem[25]{YC Acar}Y.C. Acar et al., ``Main parameters of LCxFCC
based electron-proton colliders'', arXiv:1602.03089 {[}hep-ph{]}. 

\bibitem[26]{FCC based} Y.C. Acar et al., ``FCC based ep and $\mu p$
colliders'', arXiv:1510.08284 {[}hep-ph{]}.

\bibitem[27]{Hagiwara} K. Hagiwara, D. Zeppenfeld and S. Komamiya,
``Excited lepton production at LEP and HERA'', Z. Phys. C 29, 115
(1985).

\bibitem[28]{Boudjema} F. Boudjema, A. Djouadi and J.L. Kneur, ``Excited
fermions at $e^{+}e^{-}$ and ep colliders'', Z. Phys. C 57, 425
(1993). 

\bibitem[29]{calchep} A. Belyayev, N.D. Christensen and A. Pukhov,
``CalcHEP 3.4 for collider physics within and beyond the Standard
Model'', Comput. Phys. Commun. 184 , 1729 (2013).

\bibitem[30]{cteq} D. Stump et al., ``Inclusive jet production,
parton distributions and the search for new physics'', JHEP 0310,
046 (2003).
\end{thebibliography}
\end{document}